\documentclass[%
superscriptaddress,
preprint,
showpacs,preprintnumbers,
showkeys
nobibnotes,
amsmath,amssymb,
prstab,
tightenlines,
showkeys,
floatfix,
]{revtex4-1}
\usepackage{graphicx}
\usepackage{revsymb4-1}
\usepackage{bm}
\usepackage{braket}
\usepackage{mathptmx}
\usepackage[colorlinks, linkcolor=blue, anchorcolor=blue, citecolor=blue]{hyperref}
\usepackage{hyperref}

\begin{document}

\title{Quadrupole absorption rate for atoms in circularly-polarized optical vortices} 

\author{Smail Bougouffa}
\email{sbougouffa@hotmail.com; sbougouffa@imamu.edu.sa}
\affiliation{Department of Physics, College of Science, Imam Mohammad ibn Saud Islamic University (IMSIU), P.O. Box 90950, Riyadh 11623, Saudi Arabia\\
ORCiD:  http://orcid.org/0000-0003-1884-4861}


\date{\today}

\begin{abstract}
Twisted light beams, or optical vortices, have been used to drive the circular motion of microscopic particles in optical tweezers and have been shown to generate vortices in quantum gases. Recent studies have established that electric quadrupole interactions can mediate an orbital angular momentum exchange between twisted light and the electronic degrees of freedom of atoms. Here we consider a quadrupole atomic transition mediated by a circularly-polarized optical vortex. We evaluate the transfer rate of the optical angular momentum to a $Ca^+$ ion involving the $4^2S_{1/2}\rightarrow 3^2D_{5/2}$ quadrupole transition and explain how the polarization state and the topological charge of the vortex beam determine the selection rules. 
\end{abstract}

\keywords{ Optical angular momentum transfer, quadrupole interaction , optical vortex beams, atoms-light interactions}

\pacs{ 37.10.De; 37.10.Gh }

\maketitle

\section{Introduction}\label{sec1}
The present trend towards efficiency and enhanced applications in various optical fields has led to a growing interest in developing and using twisted light beams or optical vortices \cite{TorresTorner2011, Surzhykov2015,babiker2018atoms,andrews2011structured,Yao2011a}. Laguerre-Gaussian beams have been suggested and used for numerous novel applications such as high-dimensional quantum information \cite{fickler2014interface}, quantum cryptography \cite{souza2008quantum}  and quantum memories \cite{nicolas2014quantum}.

The progress of research in laser cooling and trapping has been concentrated on the interaction of such special forms of light with atoms \cite{babiker2018atoms} involving the electric dipole interactions as well as interactions of multipolar order higher than the electric dipole \cite{ScholzMarggraf}.

In other studies quadrupole transitions have also featured and in contexts where they have been significantly enhanced \cite{tojo2005precision, hu2012v, kern2012strong}. In particular, it has been shown that the interaction of vortex light, such as Laguerre-Gaussian and Bessel-Gaussian modes, with atoms involving electric quadrupole transitions, can affect atomic motion \cite{Al-Awfi2019, Bougouffa20, Ray2020a}.

There are also several theoretical and experimental investigations that dealt with the possibility of the exchange of orbital angular momentum (OAM) between light and the internal motion of atoms \cite{van1994selection, 29, Araoka2005, loffler2012cholesteric, Giammanco2017}.  The main established result of the previous studies is the nonexistence of the OAM impact on electric dipole transitions \cite{29, Lloyd2012, lloyd2012interaction}.

Recently, the OAM transfer to atoms interacting with optical vortices was considered and the absorption rate evaluated in the cases of the $6^2S_{1/2}\rightarrow 5^2D_{5/2}$ quadrupole transition in Cs when cesium atoms are subject to the field of a linearly polarized optical vortex \cite{Bougouffa}. The obtained results showed that the absorption rate, albeit lower than the quadrupole spontaneous emission rate, is still measurable within the current experimentally available parameters. Also, it should still be within the measurement abilities of modern spectroscopic techniques.
Those studies have been mainly concerned with the case in which the optical vortex light is linearly polarized, and so optical spin is ignored in the transfer process. 
On the other hand, the experimental works by Schmiegelow et al. \cite{schmiegelow2016transfer, afanasev2018experimental}  have indicated that an atom or an ion can exchange two units of optical angular momentum, one unit from optical spin and another from its OAM. 
In this regard, the relative strengths of the corresponding transitions in the case of the target  $40 Ca^{+}$  ions have been measured  \cite{schmiegelow2016transfer, afanasev2018experimental} while the transition amplitudes have been evaluated for different twisted light ﬁelds \cite{afanasev2018experimental}.

We have thus set out to set up the necessary theory leading to the evaluation of the rate of OAM transfer from the circularly polarized optical vortices to the atoms. The theory is applied to the particular case involving the $4^2S_{1/2}\rightarrow 3^2D_{5/2}$ in $Ca^+$ ion when calcium ions are subject to the field of a circular polarized optical vortex. This $Ca^+$ transition is well known  as a dipole-forbidden but a quadrupole-allowed transition \cite{zhang2020improvement}. In evaluating the rate of OAM transfer involving a quadrupole transition we had to consider the relevant selection rules which involve both spin and OAM \cite{Rajasree2020,Klimov2009, Klimov:2012wf}.

This paper is structured as follows. In section \ref{sec2} the fundamental concepts and essential formalism involved in the interaction of twisted light with an atom in a quadrupole active transition is first outlined, leading from the quadrupole interaction Hamiltonian to the quadrupole Rabi frequency. Section \ref{sec3} is concerned with the process of OAM transfer as the atom interacts with the optical vortex field at near-resonance with the aim of evaluating the OAM transfer rate. The model treats the atom as a two-level system and applies the Fermi Golden rule with appropriate use of the selection rules for quadrupole transitions. However, the transfer process demands a treatment including the density of the continuum states as a Lorentzian function representing the upper atomic level as an energy band of width $\hbar \gamma$ where $\gamma^{-1}$ is the lifetime of the upper state. Section \ref{sec4} deals with the optical vortex as a circularly-polarised Laguerre-Gaussian beam whose winding number is restricted by the optical spin and quadrupole section rules. The results are shown in section \ref{sec5} for the quadrupole atomic transition $4^2S_{1/2}\rightarrow 3^2D_{5/2}$ in $Ca^+$ ion. Section \ref{sec6} contains a summary and final conclusions.

\section{Quadrupole interaction Hamiltonian
}\label{sec2}

We consider a two-level atom interacting with a single optical vortex beam propagating along the +z axis. The ground and excited states of the two-level-atom are $\{\ket{g}, \ket{e}\}$ with frequencies $\omega_e$ and $\omega_g$, respectively, which correspond to a transition frequency $\omega_a=(\omega_e - \omega_g)$.

We focus on the case of an optical transition that is dipole-forbidden, but quadrupole-allowed which allows us to  consider only the quadrupole interaction term in the interaction Hamiltonian, which arises from a multipolar series expansion about the center of mass coordinate  $\mathbf{R}$ as follows: 

\begin{equation}\label{2}
 \hat{H}_{Q}=-\frac{1}{2}\sum_{ij} \hat{Q}_{ij} \nabla_i \hat{E_j}.
\end{equation}
Here $x_i$ are the components of the internal position vector $\mathbf{r}=(x, y, z)$, ${\hat{Q}}_{ij}=e r_i r_j$  are the components of the quadrupole moment tensor with $r_i$ and $r_j$ the Cartesian component of the internal vector, and $\nabla_i$ are components of the gradient operator which act only on the spatial coordinates of the transverse electric field vector $\mathbf{ E}$ as a function of the centre of mass position vector variable $\mathbf{ R}= (X, Y, Z)$.
The quadrupole tensor operator  ${\hat{Q}}_{ij}$  can be written in terms of ladder operators as $ \hat{Q}_{ij}=Q_{ij}( \hat{b} + \hat{b}^{\dag})$, where $Q_{ij}=\bra{i}\hat{Q}_{ij}\ket{j}$ are the quadrupole matrix elements between the two atomic levels, and $ \hat{b} ( \hat{b}^{\dag})$ are the atomic level lowering (raising) operators.

In the following,  we assume that the electric field is circularly polarized and propagating along the $z$ direction, so optical spin will play a crucial role here, in which case we have the following form of the quadrupole interaction Hamiltonian
\begin{equation}\label{3}
  \hat{H}_{Q}=-\frac{1}{2}\sum_{i} \left[\hat{Q}_{ix}\frac{\partial  \hat{E_x}}{\partial R_i}+\hat{Q}_{iy}\frac{\partial  \hat{E_y}}{\partial R_i}\right ]
\end{equation}
The quantized electric field can conveniently be written in terms of the centre-of-mass position vector in cylindrical polar coordinates $\mathbf{R}=(\rho,\phi, Z)$ as follows
\begin{equation}\label{4}
    \mathbf{ \hat{E}}(\mathbf{R})=\left(\mathbf{ \hat{i}}\alpha+\mathbf{ \hat{j}}\beta\right) u_{\{k\}}(\mathbf{R})\hat{a}_{\{k\}}e^{i \theta_{\{k\}}(\mathbf{R})}+H.c.
\end{equation}
where the complex numbers, $\alpha$ and $\beta$, determine the polarization state of the beam, $\sigma_z$. In fact, $\sigma_z=i(\alpha\beta^*-\beta\alpha^*)$, where $\alpha$ and $\beta$ are normalized and $|\alpha|^2+|\beta|^2=1$ so that $\sigma_z=+1,0,-1$ for right circular, linear, and left circular polarizations, respectively.
Also, $u_{\{k\}}(\mathbf{R})$ and $\theta_{\{k\}}(\mathbf{R})$ are, respectively, the amplitude function and the phase function of the LG vortex electric field. Here the subscript $\{k\}$ denotes a group of indices that specify the optical mode in terms of its axial wave-vector $k$, winding number $\ell$ and radial number $p$. The operators $\hat{a}_{\{k\}}$ and $\hat{a}_{\{k\}}^{\dagger}$ are the annihilation and creation operators of the field mode $\{k\}$. Finally $H.c.$ stands for Hermitian conjugate.
Using this form of the electric field, we obtain the desired expression for the quadrupole interaction Hamiltonian 
\begin{equation}\label{5}
    \hat{H}_{Q}=\hbar\Omega^{Q}_{\{k\}}(\mathbf{R})e^{i\theta_{\{k\}}(\mathbf{R})}\hat{a}_{\{k\}}(\hat{b}^{\dag}+\hat{b})+H.c.
\end{equation}
where $\Omega^{Q}_{\{k\}}(\mathbf{R})$ is the quadrupole Rabi frequency, which can be written as
\begin{equation}\label{5p}
   \Omega^{Q}_{\{k\}}(\mathbf{R})=-\frac{1}{2\hbar}\sum_{i} \left (\alpha Q_{ix}+\beta Q_{iy}\right) u_{\{k\}}\Big( \frac{1}{ u_{\{k\}}}\frac{\partial u_{\{k\}}}{\partial R_i}+i\frac{\partial \theta_{\{k\}}}{\partial R_i} \Big)
\end{equation}

It is suitable to proceed as we show below by supposing a general LG mode LG$_{\ell p}$ of winding number $\ell$ and radial number $p$.  The values of $\ell$ and $p$ relevant to a given quadrupole transition are chosen within the selection rules of the considered atomic transition.

\section{Transition amplitude and absorption rate of OAM by atom}\label{sec3}

We consider the vortex field with an orbital angular momentum (OAM) $\pm \ell \hbar$ and a spin angular momentum (SAM) $\pm s\hbar$ per photon where $\ell$ and $s$ are positive \cite{watzel2020electrons}.
Hence, the transition matrix element \cite{Bougouffa, forbes2018chiroptical, ScholzMarggraf}, comprising only the quadrupole coupling, is specified by
$\mathrm{M}^{\{k\}}_{if}=\bra{f}\hat{H}_{Q}\ket{i}$,
where $\ket{i}$ and $\ket{f}$ are, respectively, the initial and final states of the overall quantum system (atom plus optical vortex). We assume that the system has as an initial state $\ket{i}$ with the atom in its ground state and there is one vortex photon.  The final state $\ket{f}$  consists of the excited state of the atom and there is no field mode. Thus $\ket{i}=\ket{g\{1\}_{\{k\}}}$ and $\ket{f}=\ket{e\{0\}}$.  

Using the relations 
\begin{eqnarray}\label{17ee}
\bra{\{0\}}\hat{a}^+_{\{k'\}}\ket{\{1\}_{\{k\}}}&=&0,\\ 
\bra{\{0\}}\hat{a}_{\{k'\}}\ket{\{1\}_{\{k\}}}&=&\delta_{\{k'\}\{k\}},
\end{eqnarray}
we obtain
\begin{equation}\label{8}
\mathrm{M}^{\{k\}}_{if}=\hbar\Omega^{Q}_{\{k\}}(\mathbf{R})e^{i\theta_{\{k\}}(\mathbf{R})}
\end{equation}
where  $\Omega^{Q}_{\{k\}}(\mathbf{R})$ is the quadrupole Rabi frequency. 
The final state of the system in the absorption process comprises a continuous band of energy of width $\hbar\gamma$ where $\gamma $ is the quadrupole spontaneous emission rate in free space.  In this case the absorption rate is governed by the form of Fermi's golden rule \cite{Bougouffa, Barnett2002,Lloyd2012,fox2006quantum} with a density of states 

\begin{eqnarray}\label{9}
    \Gamma_{if}= 2\pi\big |\Omega^{Q}_{\{k\}}(\mathbf{R})\big|^2\mathcal{F}_{\omega_a}(\omega),
\end{eqnarray}
where $\mathcal{F}_{\omega_a}(\omega) $ is the density of the final state and  it can be well characterized by a Lorentzian distribution of states with a width (FWHM) matching with the spontaneous quadrupole emission rate, thus
\begin{equation}\label{9ppp}
\mathcal{F}_{\omega_a}(\omega)= \frac{1}{\pi}\frac{\gamma/2}{(\omega-\omega_a)^2+(\gamma/2)^2},
\end{equation}
The Lorentzian distribution characterizing the density of states specifies a limit to the validity of using Fermi's Golden rule to calculate the absorption rate, since such a rate is valid only if the frequency width of the upper state $\ket{e}$ is larger than the excitation rate; i.e., the spontaneous emission rate is larger than the Rabi frequency.  The Rabi frequency may exceed the spontaneous emission rate for high intensities, in which case the perturbative approach culminating in the Fermi Golden Rule is no longer valid and the strong coupling regime is applicable involving Rabi oscillations.
Substituting Eq. (\ref{9ppp}) in Eq. (\ref{9}) we find for  the quadrupole absorption rate
\begin{equation}\label{main}
\Gamma_{if}=  \frac{\gamma}{(\omega-\omega_a)^2+(\gamma/2)^2}\big |\Omega^{Q}_{\{k\}}(\mathbf{R})\big|^2
\end{equation}
In the following, we are concerned with the use of the optical vortex in the Laguerre-Gaussian (LG) mode.

\section{Circularly polarized Laguerre-Gaussian Mode}\label{sec4} 

IIn the paraxial regime and using cylindrical coordinates \cite{Allen1992}, the LG$_{\ell p}$ beam is described by the amplitude function 
\begin{equation}\label{15}
    u_{\{k\}}(\rho)=u_{k\ell p}(\rho)=E_{k00}g_{\ell,p}(\rho)
\end{equation}
with
\begin{equation}\label{15p}
g_{\ell,p}(\rho)=\sqrt{\frac{p!}{(|\ell|+p)!}}\Big( \frac{\rho\sqrt{2}}{w_0}\Big)^{|\ell|}L_p^{|\ell|}(\frac{2\rho^2}{w_0^2})e^{-\rho^2/w_0^2},
\end{equation}
where $L_p^{|\ell|}$ is the associated Laguerre polynomial  and $w_0$ is the radius at beam waist  (at $Z=0$).  The global factor $E_{k00}$ is the constant amplitude of the corresponding  plane electromagnetic wave.
The phase function of the LG mode in the paraxial regime is given by
\begin{equation}\label{16}
\theta_{klp}(\rho,Z,t)\approx kZ+\ell\phi-\omega t .
\end{equation}
For the sake of simplicity, we here assume that the winding number $\ell$ of the LG beam is positive. Here, the LG$_{\ell p}$ beam is supposed to be circularly polarized and the polarisation vector $\mathbf{ \hat{\varepsilon}}=\mathbf{ \hat{i}}\alpha+\mathbf{ \hat{j}}\beta$. In brief, every photon of the beam with waist $w_0$ is in the same quantum state as categorized by the resulting four beam parameters: frequency $\omega$, radial index $p$, winding number $\ell$, and spin $\sigma_z$.

The quadrupole Rabi frequency associated with the LG$_{\ell p}$  of frequency $\omega$, which is circularly polarised in the  $x-y$ plane can be written as follows  \cite{andrews2011structured, babiker2018atoms, klimov1996quadrupole,lin2016dielectric, fickler2012quantum, Domokos2003, deng2008propagation}


\begin{equation}\label{12}
    \Omega _{k\ell p}^{Q} (\rho)=\left(u_{p}^{\ell } (\rho)/\hbar \right) \Big (G({\bf R})\mathcal{Q}_{1} +H({\bf R}) \mathcal{Q}_{2} +ik \mathcal{Q}_{3} \Big)
\end{equation}


where the modified quadrupole moments are $\mathcal{Q}_{j}=(\alpha Q_{jx}+\beta Q_{jy})$ with $j= x, y, z$, and the functions $G({\bf R})$ and $H({\bf R})$ are as follows
\begin{eqnarray}
G({\bf R}) =\left(\frac{\left|\ell \right|X}{\rho^{2} } -\frac{2X}{w_{0}^{2} } -\frac{i\ell Y}{\rho^{2} } +\frac{1}{L_{p}^{\left|\ell \right|} } \frac{\partial L_{p}^{\left|\ell \right|} }{\partial X} \right) \label{17a}\label{13},\\
H({\bf R}) =\left(\frac{\left|\ell \right|Y}{\rho^{2} } -\frac{2Y}{w_{0}^{2} } +\frac{i\ell X}{\rho^{2} } +\frac{1}{L_{p}^{\left|\ell \right|} } \frac{\partial L_{p}^{\left|\ell \right|} }{\partial Y} \right) \label{14},
\end{eqnarray}

Finally, we get the quadrupole absorption rate for an atom interacting with the LG$_{\ell,p}$ light mode that is circularly polarized in $ x-y$ plane, and the atom is characterised by the quadrupole matrix elements $Q_{xx}, Q_{xy}, Q_{yy}, Q_{zx}$ and $Q_{zy}$

\begin{eqnarray}\label{result}
\Gamma_{{if}}&=& \frac{\gamma}{(\omega-\omega_a)^2+(\gamma/2)^2}\Big| G({\bf R})\mathcal{Q}_{1} +H({\bf R}) \mathcal{Q}_{2} +ik \mathcal{Q}_{3} \Big|^2 \nonumber \\&& \times \left|u_{p}^{\ell } (\rho)/\hbar \right|^2,
\end{eqnarray}

This general result applies to any atom with a quadrupole-allowed but a dipole-forbidden transition, at near resonance with a circularly polarized Laguerre-Gaussian light mode LG $_{\ell,p}$.  The principal constraint is that the interaction must obey the OAM and SAM selection rules involving the quantum number $m$ between the ground and excited atomic states $\ket{g}$ and $\ket{e}$ and for a quadrupole transition, we have
\begin{equation}
\Delta m=0,\pm 1,\pm 2
\end{equation}
The constraint of angular momentum conservation then means that the optical vortex absorption process in a quadrupole transition can only arise for optical vortices if the total angular momentum (TAM) of a Laguerre-Gaussian beam has a value $\ell +\sigma_z$, which is not larger than the multipolarity $L$ of the underling atomic transition, i.e.,
\begin{equation}
\ell+\sigma_z \le L,
\end{equation}
where for the circularly polarized field $\sigma_z=\pm1$ and the winding numbers are  $\ell \leq L \mp 1$. The details will depend on the specific atom and its specific quadrupole transition. Note that although the radial quantum number $p$ is important for the amplitude distribution function of the LG$_{\ell,p}$ mode, the magnitude of the OAM transferred is determined exclusively by the value of the winding number $\ell$ and the spin $\sigma_z$

The case $(\ell=0, \sigma_z=0)$ is possible, but then no transfer of orbital angular momentum occurs in the absorption process, while the case $(\ell=1, \sigma_z=0)$ corresponds to transfer of total angular momentum of magnitude $\hbar$ is well investigated in \cite{Bougouffa}. Here, we will explore the cases  $(\ell \leq L \mp 1)$, which are accompanied by a transfer of total angular momentum (TAM) of magnitude $\ge \hbar$.

In the following, we will explore the main result with useful illustrations by focusing on the case that has lately been examined \cite{Bougouffa20,babiker2018atoms,lembessis2013enhanced, Al-Awfi2019, Bougouffa}, namely an LG mode of the winding number $\ell \le L\mp 1$ and the radial number $p$. In the simplest case where the mode is a doughnut mode $p=0$, we find that the last terms involving the derivatives in $U({\bf R})$ and $V({\bf R})$ given by Eqs. (\ref{13},\ref{14}) vanish, as $L_{0}^{\left|\ell \right|}$ are constants for all $\ell$. Though, the case where $p\neq 0$ is also of concern since the value of $p$ is significant for the intensity distribution. 
A particular atomic transition, we shall consider showing the results is that of the $Ca^+$ ion, namely $4^2S_{1/2}\rightarrow 3^2D_{5/2}$ transition. 

\section{Results and discussion}\label{sec5}

To explore the results, we need to evaluate the quadrupole matrix elements $Q_{xx}$, $Q_{xy}$, $Q_{yy}$, $Q_{zx}$ and $Q_{zy}$, which are related to the considered atomic transition, depending on the OAM selection rules. Indeed, using the normalized hydrogen-like wave function $\psi_{nLm}$ \cite {Bransden2003, Fischer1973}, with an appropriate value of the effective nuclear charge of $Z_a\approx 4.4$ \cite{Bransden2003,Fischer1973,Varshalovich1988,kreuter2005experimental,PhysRevLett.102.023002,zhang2020improvement},
 the quadrupole matrix elements can be calculated \cite{Bougouffa} from
$Q_{\alpha \beta} =e\bra {\psi_f} x_{\alpha}x_{\beta} \ket {\psi_i}$,
where $x_{\alpha}=(x,y,z)$. Straightforward evaluations yield the following:

\begin{itemize}
  \item for the transition $\ket {L=0,m=0}\rightarrow \ket{L=2, m=0}$, we find that  $\mathcal{Q}_{1}=\alpha Q_{xx}$, $\mathcal{Q}_{2}=\beta Q_{xx}$ and $\mathcal{Q}_{3}=0$,
  \item  for the case $\ket{L=0, m=0}\rightarrow \ket{L=2, m=\pm1} $, we have  $\mathcal{Q}_{1}=\mathcal{Q}_{2}=0$ and $\mathcal{Q}_3=i|Q_{xz}|(\alpha \pm i\beta)$,
  \item for the transition $\ket{L=0, m=0}\rightarrow \ket{L=2, m=\pm 2}$, we have $\mathcal{Q}_{2}=\pm i\mathcal{Q}_{1}=\pm i|Q_{xx}|(\alpha \pm i\beta)$ and $\mathcal{Q}_3=0$.
\end{itemize}
We consider a quadrupole  transition with the selection rules $\Delta L=2$ and  $\Delta m=0,\pm 1, \pm 2$ applicable for the $(4^2S_{1/2}\rightarrow 3^2D_{5/2})$ quadrupole transition in $Ca^+$ ion.

As the  atomic levels in question  involve both the spin of the electron  as well as the orbital angular momentum, both of which are necessary for the inclusion of spin effects in fine structure,  we need to  include the Clebsch–Gordan coefficients (CGC) in the formalism. The appropriate Clebsch–Gordan coefficients for different transitions are given as \cite{afanasev2018experimental}

\begin{eqnarray}\label{17ab}
   CGC &=&\sqrt{\frac{5}{5}}, \quad \sqrt{\frac{4}{5}}, \quad \sqrt{\frac{3}{5}}, \quad \sqrt{\frac{2}{5}}, \quad \sqrt{\frac{1}{5}},
\end{eqnarray}
which correspond to transition with $ \Delta m =-2, -1,  0, +1,  +2 .$, respectively.

\subsection{No OAM transfer case}

For this case $\Delta m=0$, expressing lengths in units of $w_0$ , so that $\bar{\rho}=\rho/w_0$ etc, the Rabi frequency  can be read as

 \begin{eqnarray}\label{17b}
    \Omega _{k\ell p}^{Q} (\rho)&=&|Q_{xx}|\left (u_{0}^{|\ell| }(\rho)/\hbar \right)\Big( \alpha G({\bf R}) +\beta H({\bf R})\Big)\nonumber\\
 &=&\Omega_{01} g_{\ell,p}(\bar{\rho})\Big ( (\frac{\ell}{\bar{\rho}^2}-2 +\frac{1}{\bar{\rho}} \frac{1}{L_{p}^{\left|\ell \right|} } \frac{\partial L_{p}^{\left|\ell \right|} }{\partial \bar{\rho}}) (\alpha \bar{X}+\beta \bar{Y})\nonumber \\
 & &+ \frac{i\ell}{\bar{\rho}^2}(\beta \bar{X}-\alpha \bar{Y})\Big)
\end{eqnarray}

where  $\Omega_{01}$ is a scaling factor for the Rabi frequency
\begin{equation}\label{19p}
\Omega_{01}=\frac{1}{\hbar}\frac{E_{k00} |Q_{xx}|}{w_0}.
\end{equation}
The requirement of OAM and SAM conservations imply that $\ell=1$ and $\sigma=-1$.
\subsection{OAM transfer case}
For this case $\Delta m=\pm 1$, the quadrupole Rabi frequency can be read as
\begin{eqnarray}\label{17c}
    \Omega _{k\ell p}^{Q} (\bar{\rho})=-\Omega_{02} (\alpha \pm i \beta) kw_0 g_{\ell,p}(\bar{\rho}),
\end{eqnarray}
where  $\Omega_{02}$ is a scaling factor for the Rabi frequency
\begin{equation}\label{19pp}
\Omega_{02}=\frac{1}{\hbar}\frac{E_{k00} |Q_{xz}|}{w_0}.
\end{equation}
The absorption rate is then given by
\begin{eqnarray}\label{20}
\Gamma_{if}&=&\frac{2\pi w_0^2}{c^2}\big | \Omega_{02} \big|^{2}\frac{\gamma/2}{\pi}\frac{\omega^2}{(\omega-\omega_a)^2+(\gamma /2)^2}\big|g_{\ell,p}(\bar{\rho})\big|^2.  \nonumber\\
\end{eqnarray}

It is clear that for $\Delta m=+1$ and the right circularly polarized field $(\alpha+i\beta=0)$, the quadrupole Rabi frequency is zero. Also, for $\Delta m=-1$ and the left circularly polarized field $(\alpha-i\beta=0)$, the quadrupole Rabi frequency is zero. Then the requirement of OAM and SAM conservations imply that $\ell=2$ and $\sigma_z=-1$ for $\Delta m=+1$.
\subsection{Total angular momentum transfer case}

In this case $\Delta m=\pm 2$, the quadrupole moments are  $\mathcal{Q}_{2}=\pm i\mathcal{Q}_{1}=\pm i|Q_{xx}|(\alpha \pm i\beta)$, and $\mathcal{Q}_3=0$ and the Rabi frequency Eq. (\ref{12}) is as follows:
\begin{eqnarray}\label{17d}
    \Omega _{k\ell p}^{Q} (\bar{\rho})&=&|Q_{xx}|(\alpha \pm i \beta)\left (u_{p}^{|\ell| }(\rho)/\hbar \right)\Big( G({\bf R}) \pm iH({\bf R})\Big),\nonumber\\
\end{eqnarray}
or 
\begin{equation}\label{17e} 
\Omega _{k\ell p}^{Q} (\bar{\rho})=\left\{ \begin{array}{ll}
\Omega_{01} g_{\ell,p}(\bar{\rho})(\alpha +i \beta)\Big (  \bar{X}+i\bar{Y}\Big)\\
 \times \Big ( -2 + \frac{1}{\bar{\rho}} \frac{1}{L_{p}^{\left|\ell \right|} } \frac{\partial L_{p}^{\left|\ell \right|} }{\partial \bar{\rho}} \Big),& \mbox{if $\Delta m= +2$};\\
\Omega_{01} g_{\ell,p}(\bar{\rho})(\alpha -i \beta)\Big (  \bar{X}-i\bar{Y}\Big) \\
\times \Big (\frac{2|\ell|}{\bar{\rho}^2} -2 + \frac{1}{\bar{\rho}} \frac{1}{L_{p}^{\left|\ell \right|} } \frac{\partial L_{p}^{\left|\ell \right|} }{\partial \bar{\rho}} \Big), & \mbox{if $\Delta m=-2$},\end{array} \right.
\end{equation}
also, for $\Delta m=+2$, the quadrupole Rabi frequency is zero for the right circularly polarized filed $\sigma_z=+1$ and inversely, the quadrupole Rabi frequency is null for the left circularly polarized field for $\Delta m=-2$. Then the requirement of OAM and SAM conservations imply that $\ell=3$ and $\sigma=-1$ for $\Delta m=+2$.

The absorption rate is then given by
\begin{equation}
\Gamma_{if}=\left\{\begin{array}{ll}
          \frac{\gamma \big |\Omega_{01} \big|^{2}\big|\bar{\rho}g_{\ell,p}(\bar{\rho})\big|^2}{(\omega-\omega_a)^2+(\gamma/2)^2}\Big|-2 + \frac{1}{\bar{\rho}} \frac{1}{L_{p}^{\left|\ell \right|} } \frac{\partial L_{p}^{\left|\ell \right|} }{\partial \bar{\rho}} \Big|^2,  &\mbox{if $\Delta m = +2$};\\
        \frac{\gamma \big | \Omega_{01} \big|^{2}\big|\bar{\rho}g_{\ell,p}(\bar{\rho})\big|^2 }{(\omega-\omega_a)^2+(\gamma/2)^2}\Big|\frac{2|\ell|}{\bar{\rho}^2}-2 + \frac{1}{\bar{\rho}} \frac{1}{L_{p}^{\left|\ell \right|} } \frac{\partial L_{p}^{\left|\ell \right|} }{\partial \bar{\rho}} \Big|^2,  &  \mbox{if $\Delta m =-2$},\end{array} \right.
\end{equation}
where  $\Omega_{01}=\frac{1}{\hbar}\frac{E_{k00} |Q_{xx}|}{w_0}$ is a scaling factor for the Rabi frequency  and $g_{\ell,p}$ is given by Eq.(\ref{15p}).

We chose the standard parameters in this case as \cite{Chan2016}   $\lambda=729 (nm)$, $|Q_{xx}|\simeq|Q_{xz}|\simeq1.85e a_0^2$ \cite{jiang2008electric},  and the spontaneous decay rate is $\Gamma_S/2\pi=2.24\times 10^7 (s^{-1})$ \cite{PhysRevLett.92.203002,kreuter2005experimental}. The beam parameters are chosen such that the beam waist $w_0=\lambda \xi$, where $\xi$ is a real number, and the intensity $I=\epsilon_0cE_{k00}^2/2$. We adopt an excessive laser intensity $I= 2\times 10^{8}Wm^{-2} $ \cite{Chan2016}, the scaling factor of the Rabi frequency can be written as
\begin{equation}\label{29b}
 \Omega_{01}\simeq\Omega_{02}\simeq \frac{1}{\hbar}\big(\frac{2I}{\epsilon_0 c}\big)^{1/2}\frac{|Q_{xx}|}{w_0}=\frac{\Omega_0}{\xi},
\end{equation} 
where $\Omega_0=2.12\times 10^{-2}\Gamma_S$.
We must make an appropriate choice of the beam waist to ensure that $\Omega_0\ll\Gamma_S$, which is the condition of the validity of the Fermi golden rule.
On the other hand, the Lorentzian density of states is chosen with a width given by the spontaneous emission rate $\gamma=\Gamma_S$, where $\Gamma_S\ll\omega_a$.

\begin{figure*}[htbp] 
\centerline{ \includegraphics[width=0.45\linewidth,height=0.31\linewidth]{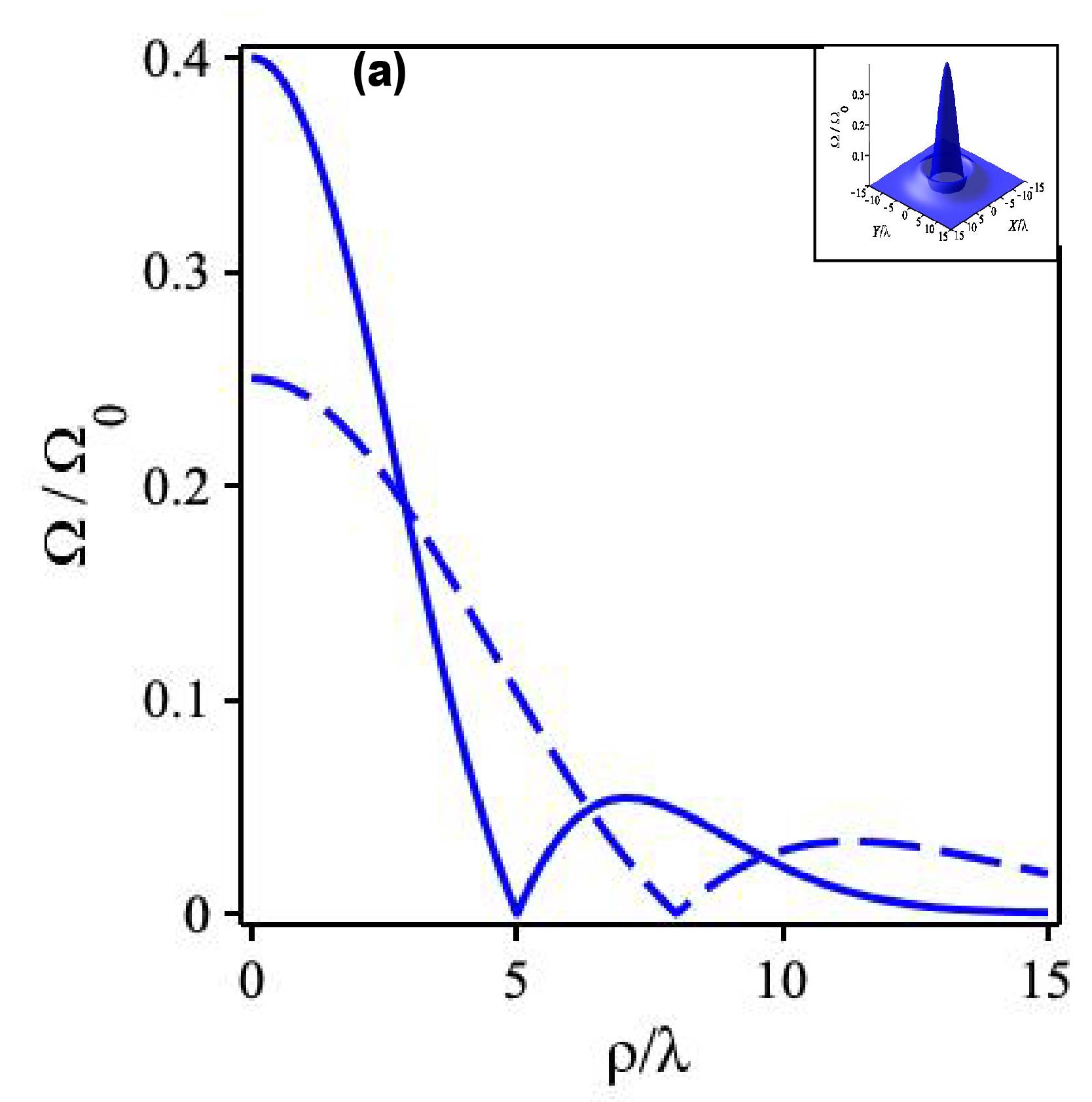}~ \includegraphics[width=0.45\linewidth,height=0.31\linewidth]{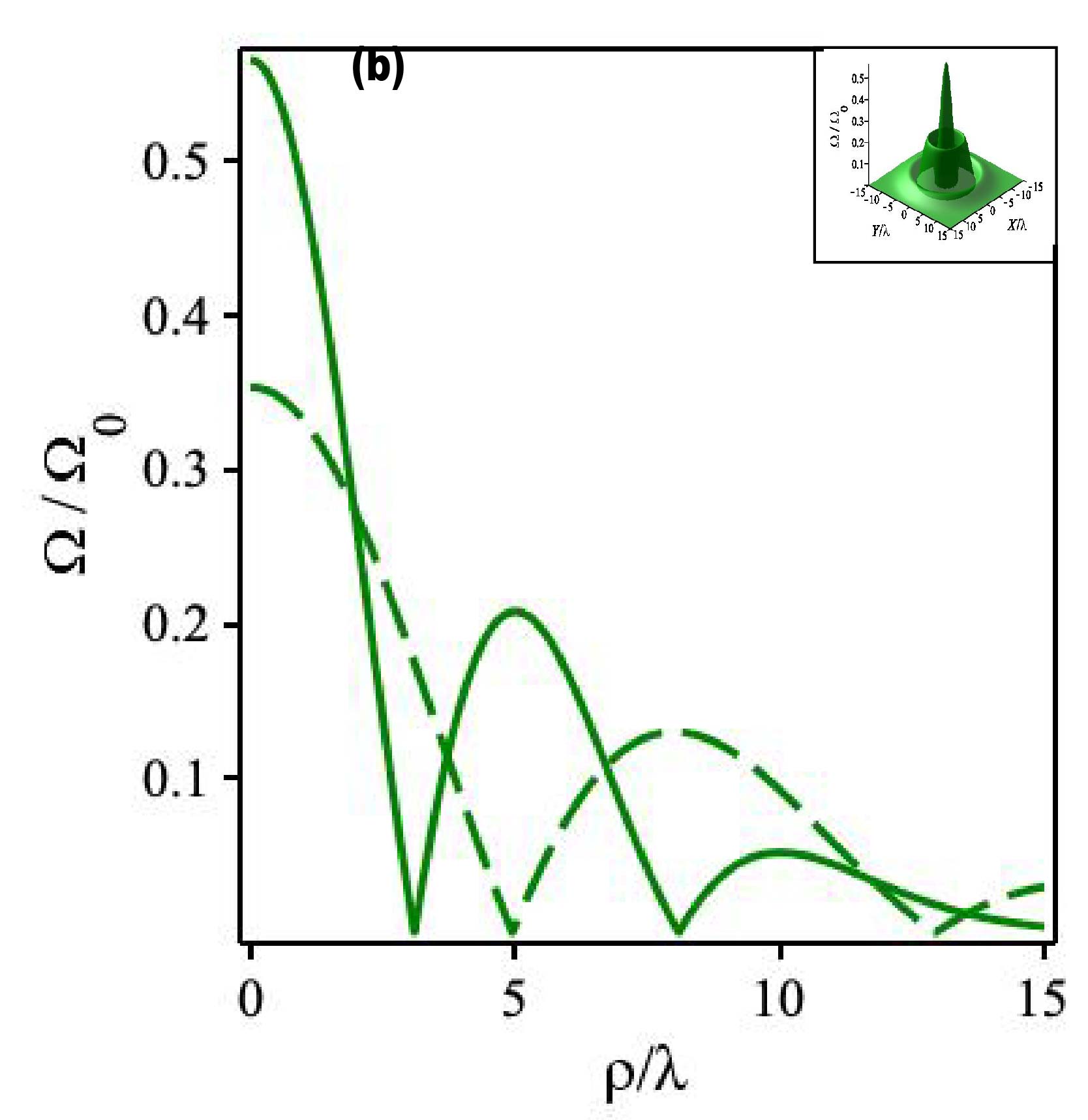} } 
\caption{(Color online) The variation with radial position of the quadrupole Rabi frequency $\Omega/\Omega_0$ for an atom in a Laguerre-Gaussian mode LG$_{\ell,p}$, $\Delta m=0$, $\ell=1$ and $\sigma_z=-1$. (a) for $p=0$. (b) for $p=1$.The solid curves are for $w_0/\lambda=5$, while the dashed curves are $w_0/\lambda=8$. The insets to the figures represent the cylindrically symmetric Rabi frequency for the case $w_0/\lambda=5$. The scaling factor $\Omega_0=\frac{1}{\hbar}\frac{E_{k00} |Q_{xx}|}{\lambda}=1.34\times 10^{-3}\Gamma_S$.}\label{Fig1} 
\end{figure*}

In Figure \ref{Fig1}, we present the variation of the quadrupole Rabi frequency  $\Omega/\Omega_0$ as a function of the radial position of the atom for different values of the beam waist $w_0/\lambda=5, 8$, and for two values of the beam radial number $p=0, 1$ for the case $\Delta m=0$, i.e., ($\ell=1, \sigma_z=-1$). It is clear that the maximum of the function  shifts away from the origin with increasing beam waist, while the value of the maximum is decreased. The insets to the figures represent the cylindrically symmetric Rabi frequency for the case $w_0/\lambda=5$. In addition, the curves are similar for the right circularly polarized field and  $\ell=-1$.

\begin{figure*}[htbp] 
\centering
 \includegraphics[width=0.45\linewidth,height=0.31\linewidth]{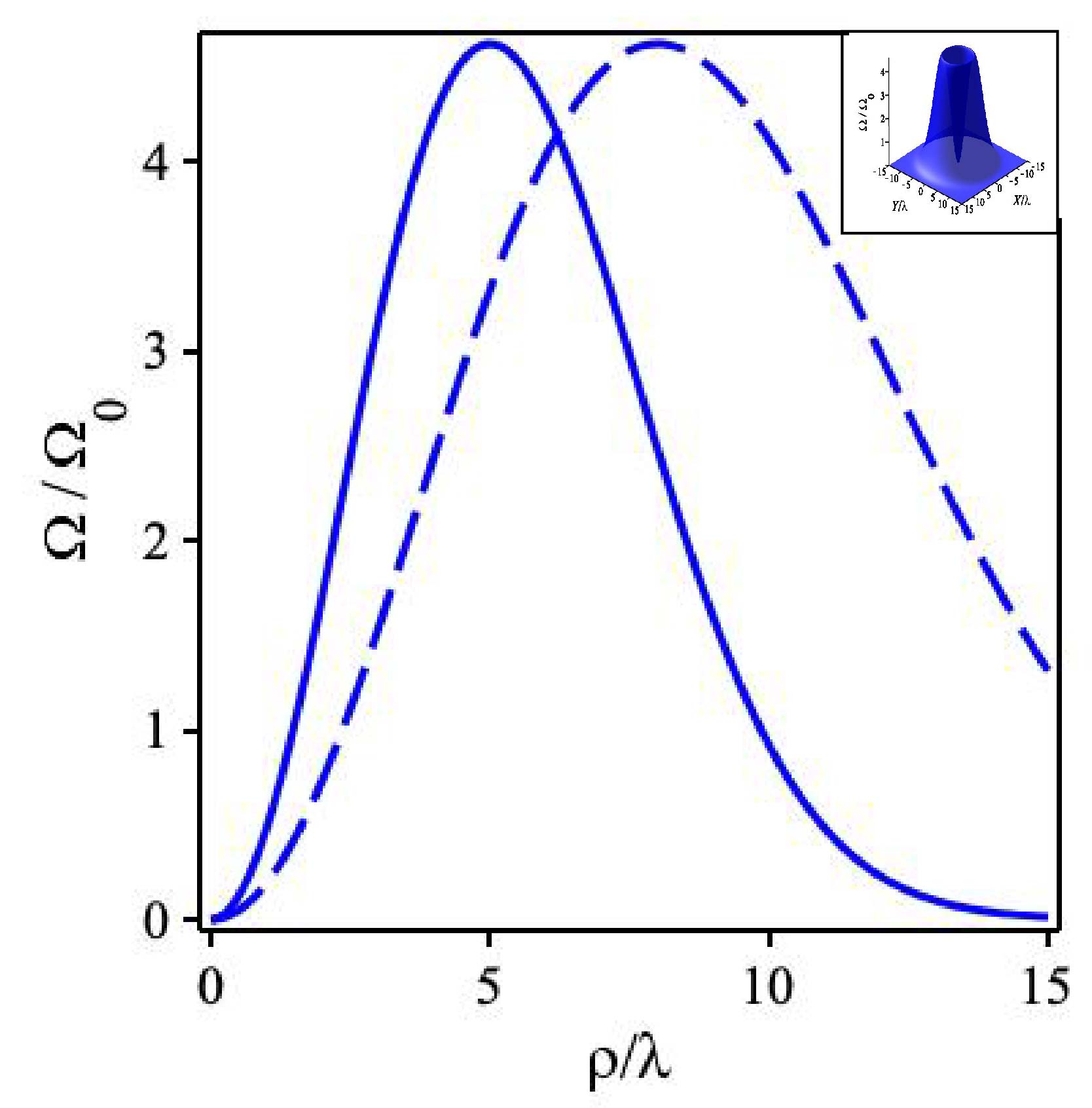}~ 
   \includegraphics[width=0.45\linewidth,height=0.31\linewidth]{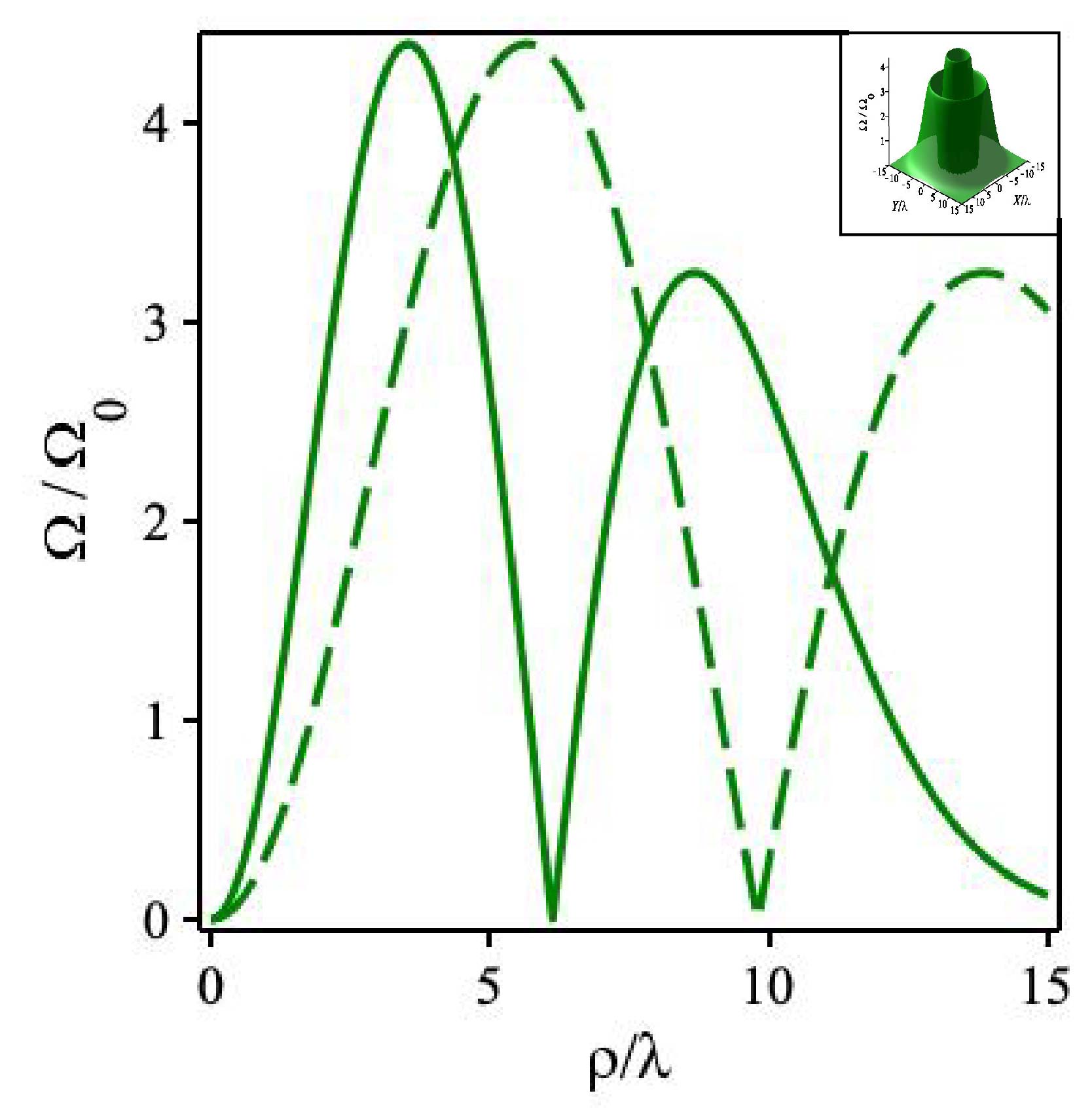} 
 \caption{(Color online) The variation with radial position of the quadrupole Rabi frequency $\Omega/\Omega_0$ for an atom in a Laguerre-Gaussian mode LG$_{\ell,p}$, $\Delta m=1$ and ($\ell=2, \sigma_z=-1$). Left panel for $p=0$. Right panel for $p=1$.The solid curves are for $w_0/\lambda=5$, while the dashed curves are $w_0/\lambda=8$.The insets to the figures represent the cylindrically symmetric Rabi frequency for the case $w_0/\lambda=5$. The scaling factor $\Omega_0=\frac{1}{\hbar}\frac{E_{k00} |Q_{xx}|}{\lambda}=1.34\times 10^{-3}\Gamma_S$.}\label{Fig2} 
\end{figure*}

In Figure \ref{Fig2}, we present the variation of the quadrupole Rabi frequency  $\Omega/\Omega_0$ as a function of the radial position of the atom for different values of the beam waist $w_0/\lambda= 5, 8$, and for two values of the beam radial number $p=0, 1$ for the case $\Delta m=+1$, i.e. $\ell=2$ and $\sigma_z=-1$. It is clear that the maximum of the function  shifts away from the origin with increasing beam waist, but the value of the maximum is independent of $w_0$. The insets to the figures represent the cylindrically symmetric Rabi frequency for the case $w_0/\lambda=5$. In addition, the quadrupole Rabi frequency is null for the right circularly polarized field. For the case $\Delta m=-1$, i.e., ($\ell=-2$ and $\sigma_z=1$), we get similar curves.

\begin{figure*}[htbp] 
\centerline{ \includegraphics[width=0.45\linewidth,height=0.31\linewidth]{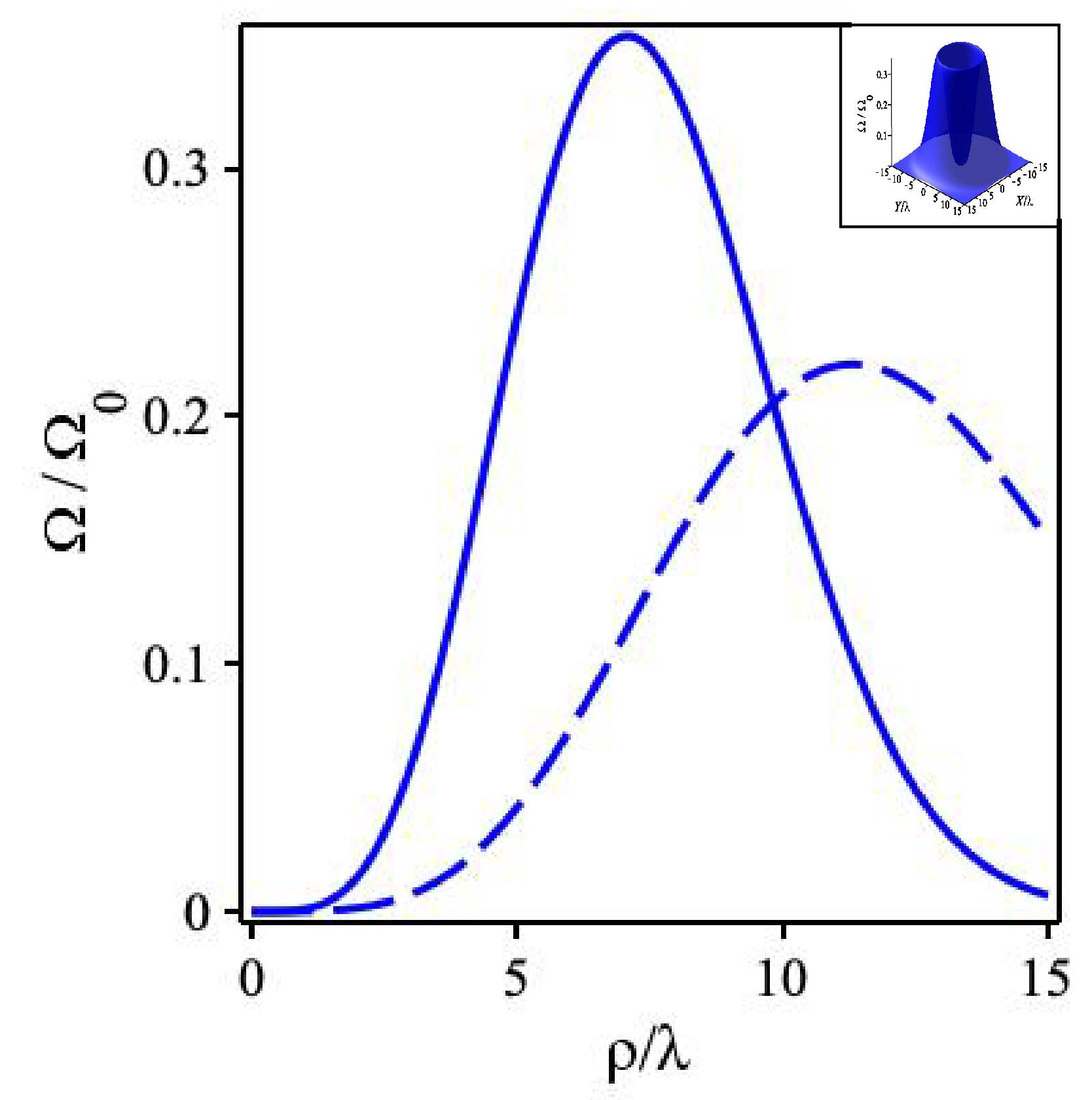}~ \includegraphics[width=0.45\linewidth,height=0.31\linewidth]{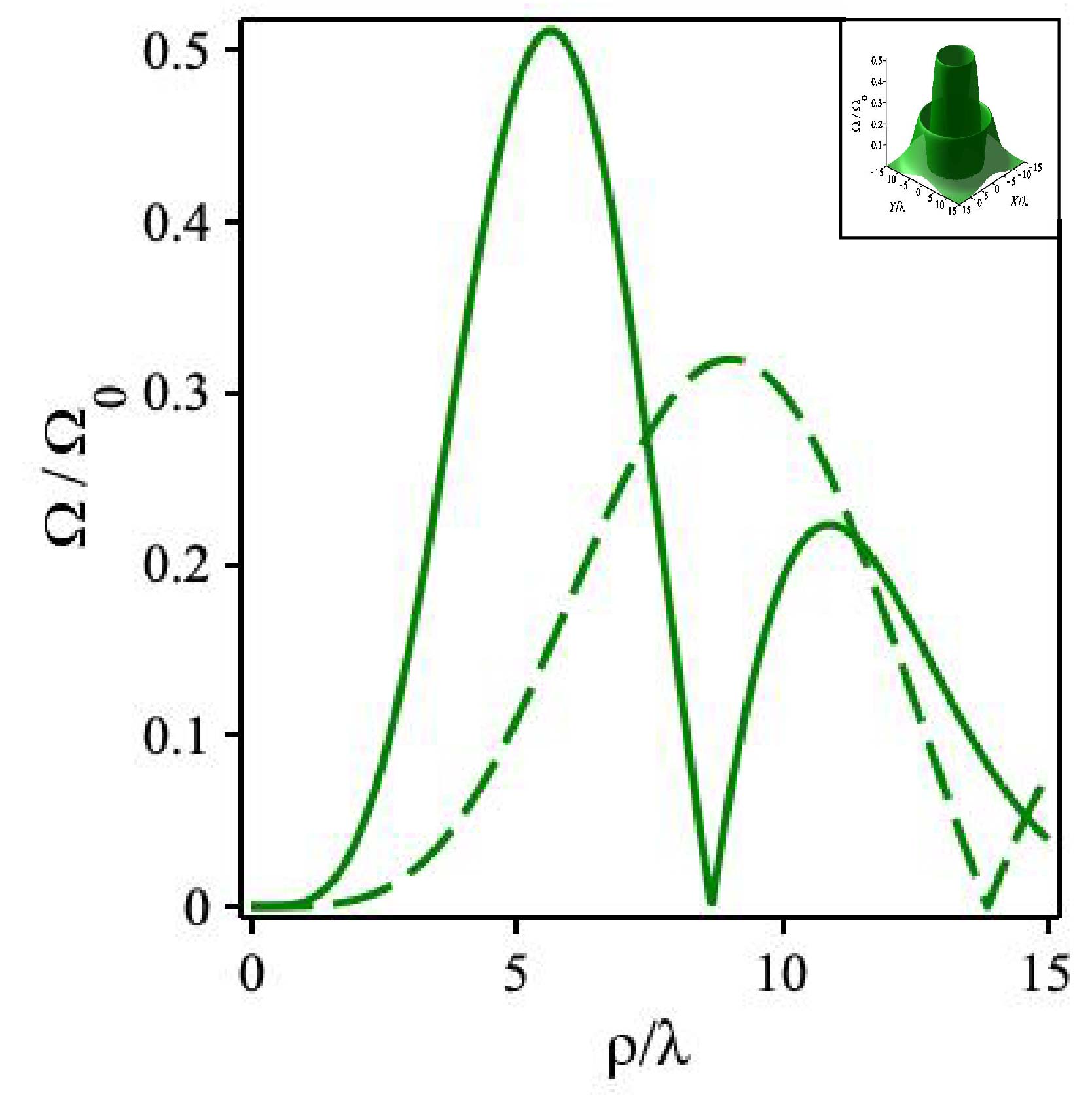} } 
\caption{(Color online) The variation with radial position of the quadrupole Rabi frequency $\Omega/\Omega_0$ for an atom in a Laguerre-Gaussian mode LG$_{\ell,p}$, $\Delta m=2$, i. e., ($\ell=3, \sigma_z=-1$). Left panel for $p=0$. Right panel for $p=1$.The solid curves are for $w_0/\lambda=5$, while the dashed curves are $w_0/\lambda=8$. The field is left circularly polarized $\sigma_z=-1$. The insets to the figures represent the cylindrically symmetric Rabi frequency for the case $w_0/\lambda=5$. The scaling factor $\Omega_0=\frac{1}{\hbar}\frac{E_{k00} |Q_{xx}|}{\lambda}=1.34\times 10^{-3}\Gamma_S$.}\label{Fig3} 
\end{figure*}

In Figure \ref{Fig3}, we present the variation of the quadrupole Rabi frequency  $\Omega/\Omega_0$ as a function of the radial position of the atom for different values of the beam waist $w_0/\lambda= 5, 8$, and for two values of the beam radial number $p=0, 1$ for the case $\Delta m=+2$, i.e., ($\ell=3, \sigma_z=-1$). It is clear that the maximum of the function  shifts away from the origin with increasing beam waist, and the maximum decreases. The insets to the figures represent the cylindrically symmetric Rabi frequency for the case $w_0/\lambda=5$. We get similar curves for the case  $\Delta m=-2$, i.e. ($\ell=-3, \sigma_z=1$).

From the figures (\ref{Fig1} , \ref{Fig3}), it is clear that the quadrupole Rabi frequency is very small comparatively to the case of the figure \ref{Fig2}, which means that the corresponding absorption rate to the case $\Delta m=+1$ is more interesting and the other cases are negligible.

\begin{figure*}[htbp]
\centerline{
 \includegraphics[width=0.48\linewidth,height=0.31\linewidth]{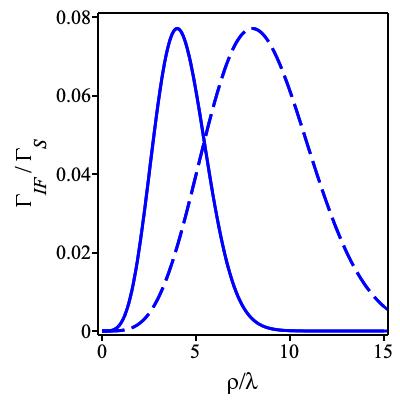}~ 
 \includegraphics[width=0.48\linewidth,height=0.31\linewidth]{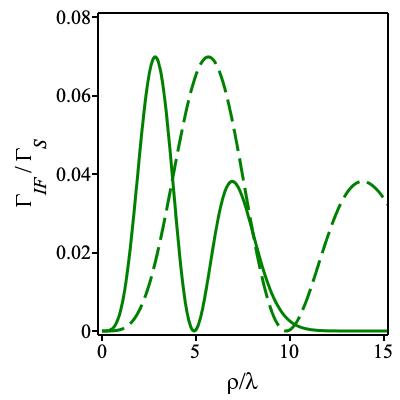}}
\caption{(Color online) The variation with radial position of the quadrupole absorption rate $\Gamma_{if}/\Gamma_S$ for $\Delta m=+1$, i.e., ($\ell=2, \sigma_z=-1$). Left panel $p=0$,  Right panel $p=1$, and an atom in a Laguerre-Gaussian mode LG$_{\ell,p}$. The solid line concerns the case $w_0/\lambda=4$; the dashed line concerns $w_0/ \lambda=8$.}\label{Fig4}
\end{figure*}

In Figure \ref{Fig4}, we present the variation of the absorption rate $\Gamma_{if}/\Gamma_S$ as a function of the radial position of the atom $\rho/\lambda$ for different values of the  beam waist $w_0/\lambda=4, 8$ for the case $\Delta m=+1$, i.e., ($\ell=2, \sigma_z=-1$). It is clear that the maximum of the function shifts away from the origin with increasing beam waist, but the value of the maximum is independent of $w_0$. 

\newpage

\section{Conclusion}\label{sec6}

In summary, we have investigated the interaction of atoms with light, which is characterized by orbital angular momentum (OAM) and spin angular momentum (SAM). 
The principal purpose is to evaluate the absorption rate of total angular momentum from the circularly polarized light to the atoms in a dipole-forbidden but quadrupole-allowed transition at near-resonance. 

We have explored a proposed technique concerning a particular case, namely the $4^2S_{1/2}\rightarrow 3^2D_{5/2}$ quadrupole transition in $Ca^+$ ion, which complies with the requirements of OAM and SAM conservation consistent with the rules $\Delta m =0,\pm 1, \pm 2$ and $\ell+\sigma_z \leq L$.

The selection rules are well established for this quadrupole transition and apply for the angular momentum of the photon to be absorbed into the final electronic state. 

We have shown that the absorption rate of orbital angular momentum by atom in quadrupole transition in the case of the $\Delta m = +1$, i.e., ($\ell=2, \sigma_z=-1$) is the dominant rate compared with other transitions. These results and analysis are in good agreement with recent experimental data as well as the results emerging from other theoretical treatments  \cite{schmiegelow2016transfer, afanasev2018experimental}. 

This work further confirms the need to continue to explore by both theory and experiment the effects of twisted light interactions with atoms. It has shown that the selection rules are motivating tools to determine the winding number of the twisted light. Furthermore, we have examined the effects of the value of the beam waist, primarily in that it shifts the position of the minima and controls the overall peak of the maxima. We have also examined the influence of the number of radial nodes and have found that the number of maxima increases with increasing radial number.  However, the first maximum is the dominant order. These predictions are worthy of experimental verification.

It is generally the case that the electric dipole transitions are dominant in the atomic spectra. But their transition rates rise with the degree of ionization.  This makes the electric dipole forbidden transitions of vital interest in the case of highly charged ions.
Also, the electric-dipole forbidden transitions are of interest for applications in plasma diagnostic and also feature in the context of astrophysics.

Furthermore, in the case of interaction of atoms with optical vortices, it is expected that the transition rates exhibit a specific impact parameter dependence, which is a common property with plasma spectroscopy \cite{SAFRONOVA200647}.

\acknowledgments {The author is grateful to Professor Mohamed Babiker for helpful discussions.}

\bibliographystyle{apsrev4-1}

\bibliography{Ref1}

\end{document}